\documentclass[aps,prb,twocolumn,10pt]{revtex4-1}
\usepackage{amsmath,amssymb,amsfonts,graphicx,times}

\begin{document}

\title{Activated scaling  in disorder rounded first-order quantum phase transitions}

\author{Arash Bellafard}
\author{Sudip Chakravarty}
\affiliation{Department of Physics and Astronomy, University of California, Los Angeles, CA 90095, USA}

\date{\today}

\begin{abstract}
First-order phase transitions, classical or quantum,  subject to randomness coupled to energy-like variables (bond randomness) can be rounded, resulting in  continuous transitions (emergent criticality). We study perhaps the simplest such model, quantum three-color Ashkin-Teller model and show that the quantum critical point in $(1+1)$ dimension  is an unusual one,  with activated scaling at the critical point and Griffiths-McCoy phase away from it. The behavior is similar to the transverse random field Ising model, even though the pure system has a first-order transition in this case. We believe that this fact must be attended to when discussing quantum critical points in numerous physical systems, which may be first-order transitions in disguise.
\end{abstract}

\maketitle

\section{Introduction}

The effect of quenched randomness on  thermodynamic properties could be varied. The systems that behave less and less random at larger and larger length scales, i.e., the randomness averages out,  are described by pure fixed points. On the other hand, if the randomness is competitive at all scales, the system is controlled by random fixed point and the properties of the system is altered by rare spatially localized active regions.~\cite{harrisJPC1974,chayesPRL1986,motrunichPRB2000} In the extreme limit, the fixed point is captured by the infinite randomness fixed point: the  main features are a strong dynamical anisotropy and a broad distribution of physical quantities which is manifest through drastically different average and typical correlation functions. Some example of such systems are the quantum critical point of random quantum Ising and Potts models,~\cite{shankarPRB1987,fisherPRL1992,senthilPRL1996,hymanPRL1997} the random singlet states of certain random antiferromagnetic spin chains,~\cite{maPRL1979,dasguptaPRB1980,mccoyPR1968I,mccoyPR1969II,mccoyPR1969III} quantum critical points separating random singlet states and the Ising antiferromagnetic phase, or the Haldane state in the random spin-1 Heisenberg chain.~\cite{monthusPRL1997}

In addition to the singularities of the thermodynamic quantities at the quantum critical point, there is a whole parameter range around the phase transition point in which physical observables display singular and even divergent behavior in spite of a finite correlation length.~\cite{fisherPRL1992,fisherPRB1995,riegerPRB1996,youngPRB1996,guoPRL1994} Within this Griffiths-McCoy phase, there is a continuously varying dynamical exponent, $z$, that relates the scale of energy and length via $\varepsilon \propto \xi^{-z},$ with $z$ diverging as $z \propto \delta^{-\psi \nu}.$ Here, $\delta$ is the deviation from the critical point, $\psi$ is some dimensionless positive constant, and $\nu$ is the correlation length  exponent. A signature of the existence of infinite randomness fixed point is the divergence of the dynamical critical exponent $z$ at the critical point, $\delta = 0$. In that case, the system exhibits  \emph{activated} dynamical scaling, $\xi_\tau \propto e^{\rm{const} \times \xi^{\psi}}$, where $\xi_\tau$ represents a characteristic time scale of the system.

Both quantum and classical \emph{first-order} phase transitions are ubiquitous in nature, because they do not require fine tuning of a control parameter of the system. Understanding the effect of quenched randomness that couples to energy-like variables on the thermodynamic properties of the systems that exhibit a first-order phase transition has been a challenge of  experimental and theoretical studies for many years.~\cite{goswamiPRL2008,greenblattPRL2009,greenblattPA2010,hrahshehPRB2012,bellafardPRL2012,bellafardAnnals2015,barghathiPS2015,imryPRB1979,huiPRL1989,aizenmanCMP1990,aizenmanPRL1989}

Here we investigate the effect of quenched disorder on the {\em quantum} three-color Ashkin-Teller model in $(1+1)$ dimension, which exhibits a first-order quantum phase transition in the absence of impurities. We employ discrete-time quantum Monte-Carlo method.  Because there is no frustration in this system, we are able to use highly efficient cluster algorithms.~\cite{bellafardPRL2012} For this disorder rounded quantum critical point, we  find activated scaling at criticality and the off-critical region is characterized by Griffiths-McCoy singularities.

The outline of this paper is as follows: in the next section, we introduce the $N$-color quantum Ashkin-Teller model. In Sec.~\ref{sec:cp}, we explain how we find the critical point. We show the evidence for activated scaling in Sec.~\ref{sec:fss}. Our results for correlation function and local susceptibility are presented in Sec.~\ref{sec:corr_func} and Sec.~\ref{sec:loc_susc}. Lastly, in Sec.~\ref{sec:dis} we provide a discussion of our findings.

\section{The Model}\label{sec:model}

The Hamiltonian of the $N$-color quantum Ashkin-Teller model in $(1+1)$ dimension is given by~\cite{goswamiPRL2008}
\begin{equation}\label{eq:qat_hamiltonian}
\begin{aligned}
	H = &-\sum_{\alpha = 1}^N \sum_{i = 1}^{L} (J_{2,i} \sigma_{\alpha,i}^{z} \sigma_{\alpha,i+1}^{z} + h_{1,i} \sigma_{\alpha,i}^{x})\\
	&- \sum_{\alpha<\beta}^N \sum_{i=1}^L (J_{4,i} \sigma_{\alpha,i}^{z} \sigma_{\alpha,i+1}^{z} \sigma_{\beta,i}^{z} \sigma_{\beta,i+1}^{z} + h_{2,i} \sigma_{\alpha,i}^{x} \sigma_{\beta,i}^{x}),
\end{aligned}
\end{equation}
where $L$ is the length of the lattice, Greek sub-indices denote the colors, Latin sub-indices denote the lattice sites, and $\sigma$'s are the Pauli operators. The $J_{2,i}$ and $J_{4,i}$ are the random nearest-neighbor coupling constants. The $h_{1,i}$ and $h_{2,i}$ are the random transverse fields. The random coupling constants and the transverse fields are taken from a distribution restricted to only positive values. The model is self-dual, which amounts to the invariance of the Hamiltonian in Eq.~\eqref{eq:qat_hamiltonian} under the transformation
$J_{2,i} \leftrightarrow h_{1,i}$,
$J_{4,i} \leftrightarrow h_{2,i}$,
$\mu_{\alpha,i}^{x} \leftrightarrow \sigma_{\alpha,i}^{z} \sigma_{\alpha,i+1}^{z}$,
and $\sigma_{\alpha,i}^{x} \leftrightarrow \mu_{\alpha,i}^{z} \mu_{\alpha,i+1}^{z}$,
where $\mu$'s are the dual Pauli operators. The pure version of this  model has been  studied in the past. It is known that for $N \ge 3$, $J_{4,i}/J_{2,i} > 0$ and $h_{2,i}/h_{1,i} > 0$, there is a  first-order phase transition from a paramagnetic to an ordered state.~\cite{grestPRB1981,fradkinPRL1984,shankarPRL1985,ceccattoJPA1991}

To study the $d$-dimensional quantum Hamiltonian in Eq.~\eqref{eq:qat_hamiltonian}, we propose an effective classical model in $(1+1)$ dimension, where the extra imaginary time dimension is of size $\beta \equiv 1/T$ and is divided up into $L_\tau \equiv \beta/\Delta \tau$ intervals each of width $\Delta \tau$ in the limit $\Delta \tau \to 0$. We introduce disorder only in the horizontal direction.  This emulates a quenched disordered quantum system whose disorder is perfectly correlated in the imaginary time direction. Hence, we expect the behavior of this system to be in the universality class as the original quantum Ashkin-Teller model in Eq.~\eqref{eq:qat_hamiltonian}. This procedure is the same as  the McCoy-Wu random Ising model,~\cite{mccoyPR1968I,mccoyPR1969II,mccoyPR1969III,shankarPRB1987} which is shown to be equivalent to the random transverse field quantum spin-$\frac{1}{2}$ Ising model in the large imaginary time limit.

The partition function is
$Z = \lim_{\Delta \tau \to 0} \mathrm{Tr}~e^{-\mathcal{S}}$,
with the proposed effective  action given by
\begin{equation}\label{eq:action}
\begin{aligned}
	\mathcal{S} =
	&- \sum_{\alpha,\tau,i} J_{i} S_{\alpha,i}(\tau) S_{\alpha,i+1}(\tau) \\
	&- \sum_{\alpha,\tau,i} J S_{\alpha,i}(\tau) S_{\alpha,i}(\tau+1)\\
	&- \sum_{\alpha \neq \beta,i,\tau} K_{i} S_{\alpha,i}(\tau) S_{\beta,i}(\tau) S_{\alpha,i+1}(\tau) S_{\beta,i+1}(\tau)\\
	&- \sum_{\alpha \neq \beta,i,\tau} K S_{\alpha,i}(\tau) S_{\alpha,i}(\tau+1) S_{\beta,i}(\tau) S_{\beta,i}(\tau+1),
\end{aligned}
\end{equation}
where the $S_i(\tau) = \pm 1$ are classical Ising spins, the indices $\alpha$ and $\beta$ denote the colors, the index $i$ runs over the sites of the  one-dimensional lattice, and $\tau = 1, 2, \dots, L_\tau$ denotes a time slice.  For computational convenience, we set $\Delta \tau = 1$ and equivalently take the limit $L_\tau \to \infty$ implying $T \to 0$. The two- and four-spin couplings, $J_i$ and $K_i$, are independent of $\tau$, because they are quenched random variables. We independently take the couplings $J_i$ and $K_i$ from the following rectangular distributions
\begin{equation}
\begin{aligned}
	\pi(J_i) &= \begin{cases}
		1, &\text{ if }\quad J-\frac{\Delta_J}{2} < J_i < J+\frac{\Delta_J}{2}\\
		0, &\text{ otherwise}
	\end{cases}\\
	\rho(K_i) &= \begin{cases}
		1, &\text{ if }\quad K-\frac{\Delta_K}{2} < K_i < K+\frac{\Delta_K}{2}\\
		0, &\text{ otherwise }
	\end{cases}
\end{aligned}
\end{equation}

Suppose we keep one of the colors in Eq.~\eqref{eq:action} fixed, for instance $\alpha = 1$. Then, we can write the Eq.~\eqref{eq:action} as
\begin{equation}\label{eq:modaction}
\begin{aligned}
	&\mathcal S = \mathcal S_{\overline 1} \\
	&- \sum_{\tau,i} \left( J_{i} + \sum_{\beta\neq 1} K_i S_{\beta,i}(\tau) S_{\beta,i+1}(\tau) \right) S_{1,i}(\tau) S_{1,i+1}(\tau)\\
	&- \sum_{\tau,i} \left( J + \sum_{\beta\neq 1} K S_{\beta,i}(\tau) S_{\beta,i}(\tau+1) \right) S_{1,i}(\tau) S_{1,i}(\tau+1),
\end{aligned}
\end{equation}
where the first term, $\mathcal S_{\overline 1}$, does not contain the color 1. The second and third terms of the Eq.~\eqref{eq:modaction} can be regarded as the Ising model action with coupling constants $J_{i} + \sum_{\beta\neq 1} K_i S_{\beta,i}(\tau) S_{\beta,i+1}(\tau)$ in the spatial direction and $J + \sum_{\beta\neq 1} K S_{\beta,i}(\tau) S_{\beta,i}(\tau+1)$ in the temporal direction. We can implement any cluster Monte-Carlo algorithm suited for the Ising model. We use the generalization of the Swendsen-Wang~\cite{swendsenPRL1987} cluster Monte-Carlo algorithm suggested by Niedermayer.~\cite{niedermayerPRL1988}

In our simulation on a square lattice of size $L \times L_\tau$ we use periodic boundary conditions in both spatial and imaginary time directions. The equilibration ``time'' is estimated using the logarithmic binning method, i.e., we compare the average values of each observable over $2^n$ Monte-Carlo steps and make sure that the last three averages are within each others error bars. Each observable is obtained by averaging over $10000$ disordered configurations and for each disordered configuration, $10000$ thermal averages is conducted. The error bars are calculated using the Jacknife procedure.~\cite{young2015everything,youngARX2012,wuAOS1986}

\section{Critical Point}\label{sec:cp}

We estimate the location of the quantum critical point along the analysis of~\citeauthor{riegerPRL1994}~\cite{riegerPRL1994} for the quantum spin glass systems using  the magnetic Binder cumulant~\cite{binderPRB1984,*binderPRB1986}
\begin{equation}\label{eq:bindercumulant}
	V_m = 1 - \frac{[\langle m^4 \rangle]}{3[\langle m^2 \rangle^2]},
\end{equation}
where
\begin{equation}
	m = \frac{1}{L_\tau L} \left[ \left\langle \sum_\alpha |m_\alpha| \right\rangle \right],
\end{equation}
with $m_\alpha = \sum_{\tau,i} S_{\alpha,i}(\tau)$. The square and angular brackets, $[\cdots]$ and $\langle \cdots \rangle$, denote the disorder and thermal averages, respectively. In the disordered phase, $V_m \propto L^{-d} \to 0$ as $L \to \infty$.~\cite{binderPRL1981,binderZPB1981} In the ordered phase, we have spontaneous magnetization at $\pm m$ and $V_m \to 2/3$ as $L \to \infty$.~\cite{binderPRL1981,binderZPB1981} Furthermore, in the paramagnetic phase, for small $L_\tau$, the system is disordered and effectively classical at a finite temperature, therefore $V_m \to 0$. For $L_\tau \to \infty$, the system is quasi one-dimensional in the imaginary time direction, therefore $V_m \to 0$ also. There exists an intermediate point where $V_m$ acquires a maximum value $V_m^\text{max}$. This maximum value decreases as $L$ increases if the system is in the paramagnetic phase, whereas it increases as $L$ increases if the system is in the ferromagnetic phase. 
There is an intermediate point at which the $V_m^\text{max}$ is a constant for all $L$ which is the quantum critical point; see Fig.~\ref{fig:atvmscaledconv}A. For our model with the parameter set $(K, \Delta_K, \Delta_J) = (0.08, 0.04, 0.2)$, we estimate the critical point to be $J_c = 0.245 \pm 0.001$.

We also found the critical point of the system with the parameter set $(K, \Delta_K, \Delta_J) = (0.1, 0.05, 0.2)$, with $J_c = 0.205 \pm 0.002$. Careful analyses of two parameter sets $(K, \Delta_K, \Delta_J) = (0.08, 0.04, 0.2)$ and $(0.1, 0.05, 0.2)$ yielded very similar results. Henceforth, we will be reporting only on the former parameter set in the rest of our paper.

\section{Finite-Size Scaling}\label{sec:fss}

The Binder cumulant~\eqref{eq:bindercumulant} has the finite-size scaling form~\cite{binderPRB1984}
\begin{equation}\label{eq:vm_scaling_form}
	V_m = \mathcal V\left(\frac{J-J_c}{J_c} L^{1/\nu},\frac{L_\tau}{L^z}\right).
\end{equation}

\begin{figure}[tbp]
	\begin{center}
		\includegraphics[width=\linewidth]{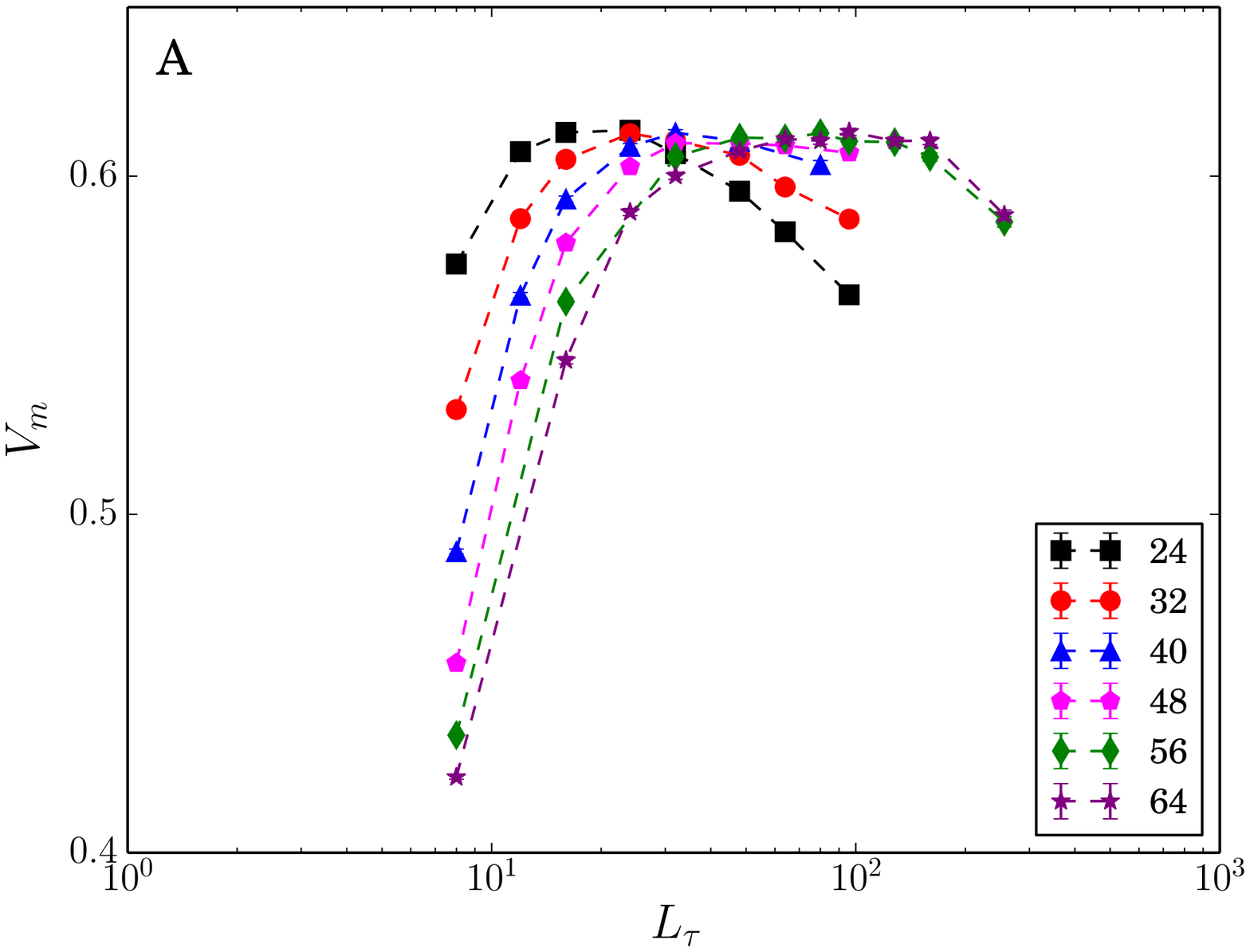}
		\includegraphics[width=\linewidth]{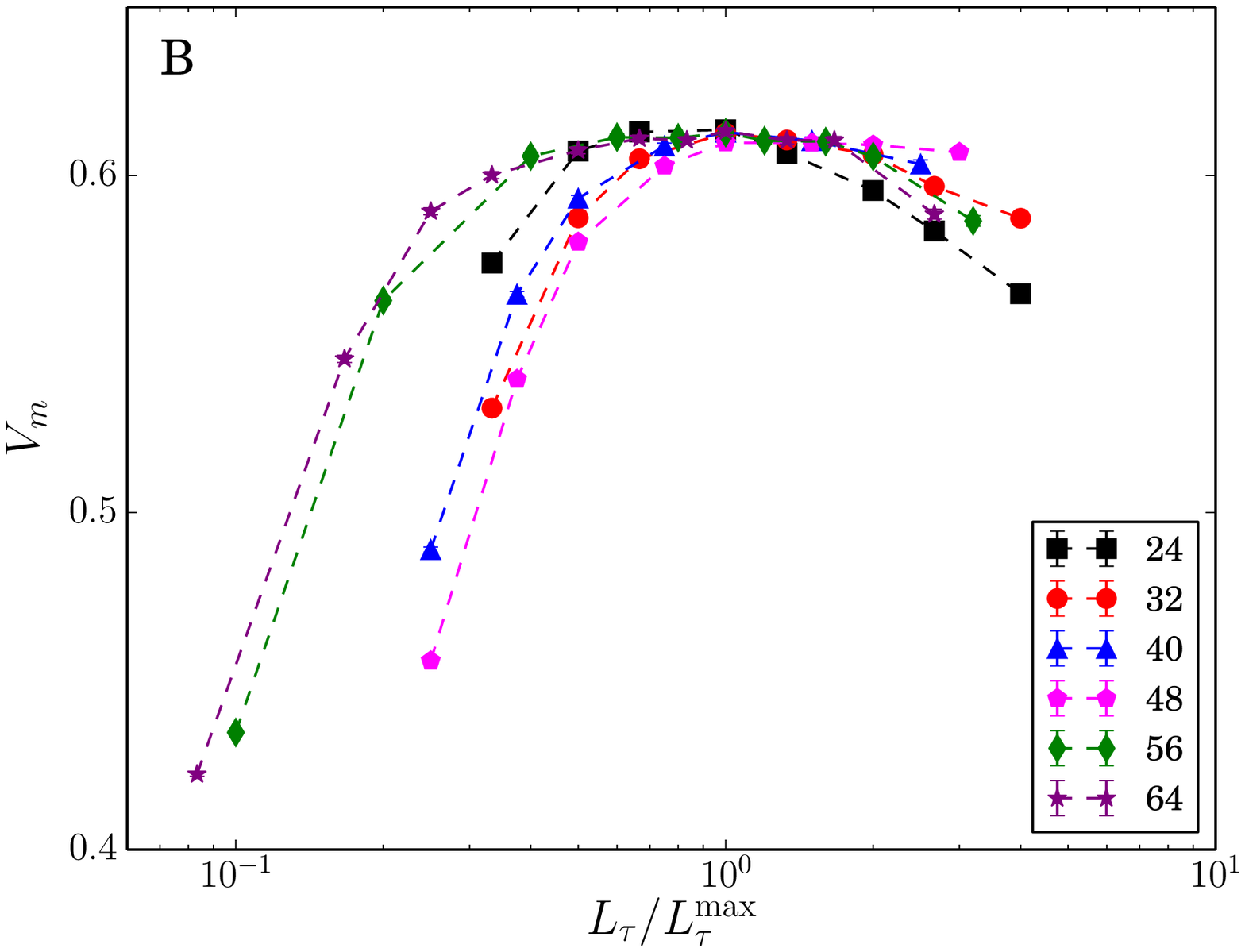}
		\includegraphics[width=\linewidth]{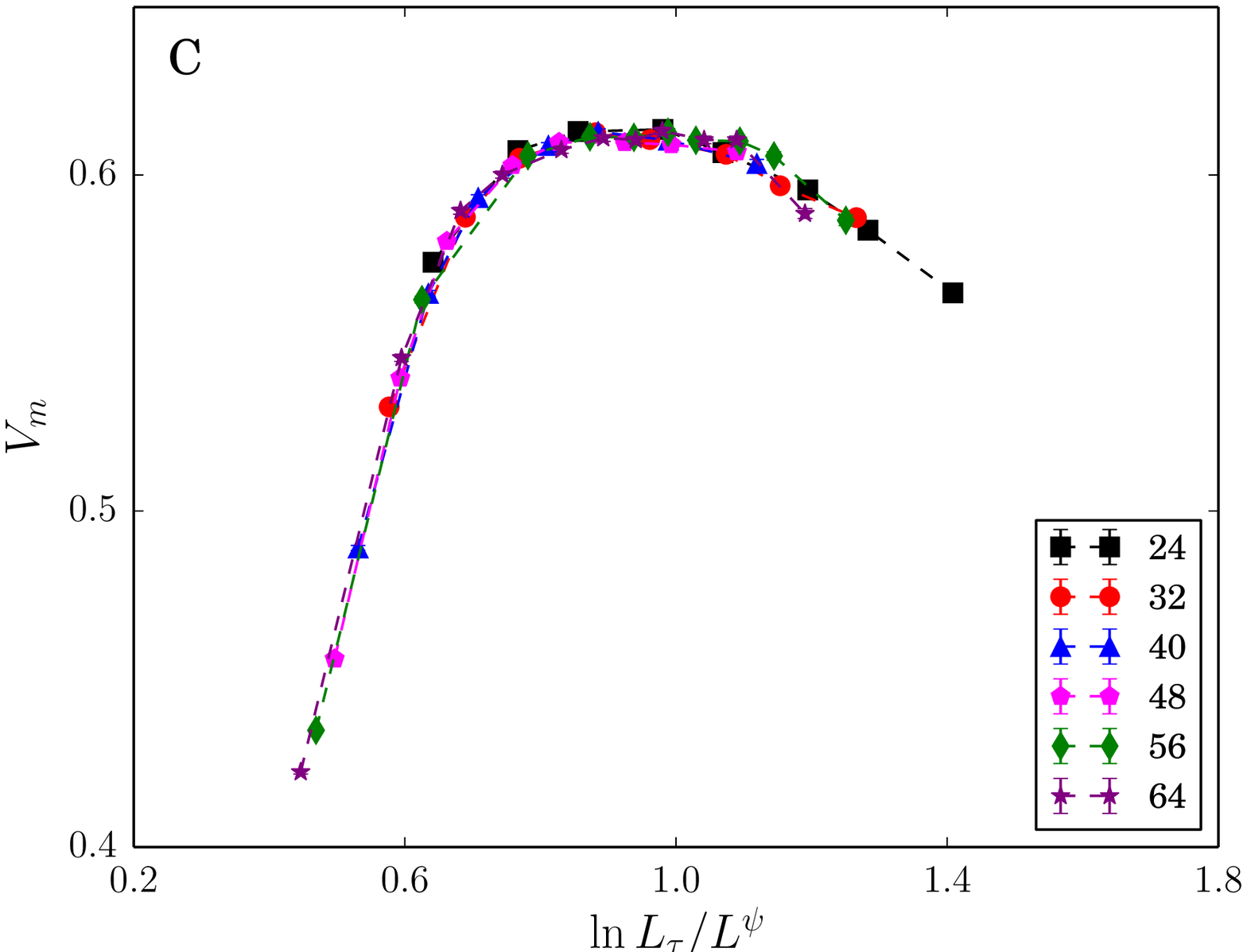}
	\end{center}
	\caption{(Color online) Magnetic Binder cumulant $V_m$ for the parameter set $(K, \Delta_K, \Delta_J) = (0.08, 0.04, 0.2)$ at $J = J_c = 0.245$. Top (A): $V_m^{\rm max}$ is $L$ independent indicating that the system is at the critical point. Middle (B): the horizontal axis is $L_\tau/L_\tau^{\rm max}$, $L_\tau^{\rm max}$ is the value of the $L_\tau$ at the peak. The curves do not scale but get broader for larger system sizes, indicating activated scaling. Bottom (C): $V_m$ versus $\ln L_\tau/L^{\psi}$ with $\psi = 0.37$. The curves scale well and is consistent with activated scaling. The actual value of $\psi$ is quite uncertain, however, and can range between $0.3-0.5$.}
	\label{fig:atvmscaledconv}
\end{figure}

As shown in Fig.~\ref{fig:atvmscaledconv}A, the value of $V_m^{\rm max}$ at the critical point is independent of the system size $L$ and $L_\tau$ at the maximum varies as $L^z$. Therefore, we naively would expect that a plot of the $V_m$ against $L_\tau/L_\tau^{\rm max}$ at the critical point should collapse the data, but from Fig.~\ref{fig:atvmscaledconv}B we see that it does not. In contrast, if we assume that the logarithm of the characteristic time scale is a power of the length scale, as in the quantum spin-$\frac{1}{2}$ Ising chain, the scaling variable should be $\ln L_\tau/\ln L_\tau^{\rm max}$ with $\ln L_\tau^{\rm max} \propto L^\psi$, for some positive constant $\psi$. As shown in the bottom of Fig.~\ref{fig:atvmscaledconv}C, the data do collapse well for $\psi = 0.37$.

\section{Correlation Function}\label{sec:corr_func}

\begin{figure}[tbp]
 	\begin{center}
 		\includegraphics[width=\linewidth]{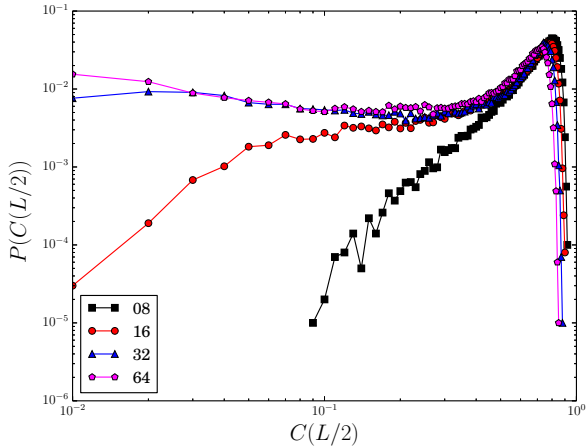}
 	\end{center}
 	\caption{(Color online) A plot of the distribution of the equal-time correlation of spins $L/2$ apart for the parameter set $(K, \Delta_K, \Delta_J) = (0.08, 0.04, 0.2)$ at $J_c = 0.245$. One sees that the distribution gets broader and broader as $L$ increases. For this plot we used $10^{5}$ realizations of disorder. The values of $L_\tau$ are chosen such that $V_m \approx V_m^{\rm max}$, namely, $L\times L_\tau \in \{8\times 9, 16\times 16, 32\times 37, 64\times 106\}$.}
 	\label{fig:atxcordist}
\end{figure}

The equal time correlation function,
\begin{equation}
	C_{\alpha,i}(r) = [\langle S_{\alpha,i}(\tau) S_{\alpha,i+r}(\tau) \rangle],
\end{equation}
is  calculated at criticality for  spins $r = L/2$ apart. As shown in Fig.~\ref{fig:atxcordist}, the distribution of the correlation function, $P(C(L/2))$, is getting broader and broader as $L$ increases. This indicates that the rare events dominate the critical properties of the system.

\begin{figure}[tbp]
	\begin{center}
		\includegraphics[width=\linewidth]{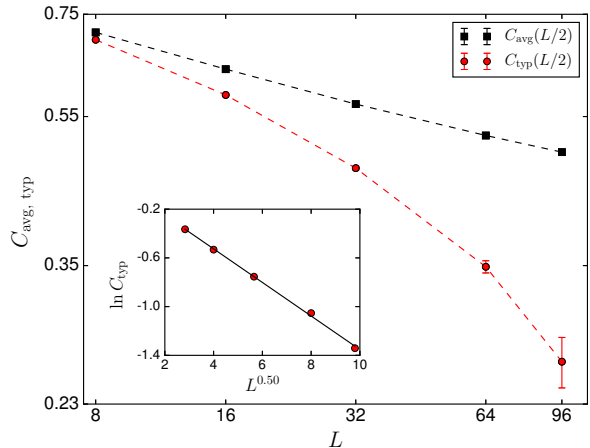}
	\end{center}
	\caption{(Color online) Average and typical correlations between spins $L/2$ apart \emph{at criticality}, $J_c = 0.245$, for the parameter set $(K, \Delta_K, \Delta_J) = (0.08, 0.04, 0.2)$ (see Fig.~\ref{fig:atxcordist}. Number of disorder realization for the size $L\times L_\tau = 96\times 224$ is $25\times 10^3$). The average falls off with a power law. The slope of the average correlation function data suggests that $\eta \approx 0.15$. The curvature of the data for the typical correlation function shows that this falls off faster than a power law. The inset shows the linear fit of the logarithm of the typical correlation function against $L^\sigma$ for the value of $\sigma = 0.50$.}
	\label{fig:atxcor}
\end{figure}

As a result of the breadth of the distribution, the \emph{average} and \emph{typical} quantities behave differently. The typical correlation function is defined here as the exponential of the average of the logarithm.~\cite{kiskerPRB1998} In Fig.~\ref{fig:atxcor}, we show that the average correlation function, $C_{\rm avg}(L/2)$, falls off as a power law,
$C_{\rm avg}(r) \propto r^{-\eta}$,
whereas the typical correlation, $C_{\rm typ}(L/2)$, has a downward curvature and falls off faster than the average value. Our result is consistent with the existence of a stretched exponential decay,
$C_{\rm typ}(r) \propto e^{-\text{const}\times r^\sigma}$,
at the critical point.

\section{Local susceptibility}\label{sec:loc_susc}

\begin{figure}[tbp]
	\begin{center}
		\includegraphics[width=\linewidth]{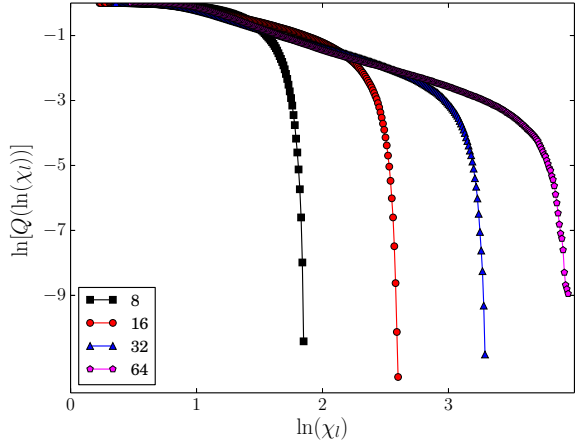}
	\end{center}
	\caption{(Color online) Cumulative probability distribution of $\ln(\chi_l)$ for the parameter set $(K, \Delta_K, \Delta_J) = (0.08, 0.04, 0.2)$ at $J = 0.232$. The distributions get broader as $L$ increases. The slope, $-d/z$, is extracted by performing a linear fit to the linear part of the largest calculated system size, namely for $L = 64$ within the region $1 \leq \ln\chi_l \leq 2.5$.}
	\label{fig:atchidistcum}
\end{figure}

We now turn our attention to \emph{off-critical} region and calculate the linear susceptibility, $\chi_l$, in the disordered phase, $J < J_c$. In the imaginary time formalism~\cite{riegerPRL1994}
\begin{equation}
	\chi_l = \sum_{\tau = 1}^{L_\tau} \langle S_{\alpha,i}(0) S_{\alpha,i}(\tau) \rangle.
\end{equation}
The dynamical exponent, $z$, can be calculated from the probability distribution of linear local susceptibility. Away from the critical point the distributions for different system sizes are well localized. Close to the critical point, however, the probability distribution of $\ln \chi_{l}$ gets broader with $L$ as shown in Fig.~\ref{fig:atchidistcum}. This broadening of the probability distribution is a strong support for the existence of strongly coupled rare regions in the vicinity of the critical point.

We examine the behavior of the distribution of local susceptibility following Refs.~\onlinecite{riegerPRB1996,youngPRB1996,guoPRL1994,igloiPRB1999}. Given that the probability distribution of logarithm of local susceptibility $P(\ln\chi_l)$ has a power law tail with $P(\ln\chi_l) \propto \chi_l^{-d/z}$, then its integral, $Q(\ln\chi_l) = \int_{\ln\chi_l}^\infty P(\ln\chi'_l) d\ln\chi'_l$, behaves similarly to $P(\ln\chi_l)$ with~\cite{riegerPRL1994}
\begin{equation}
	\ln[Q(\ln\chi_l)] = -\frac{d}{z}\ln\chi_l + \text{const}.
\end{equation}
It is more accurate to extract the  exponent, $z$, from the cumulative distribution, $Q(\ln(\chi_l))$. In Fig.~\ref{fig:atchidistcum}, we show the cumulative distribution of the logarithm of local linear susceptibility. 

From the conservation of the probability distribution, we have $\int P(\ln\chi_l) d\ln\chi_l = \int \tilde P(\chi_l) d\chi_l$. Therefore $\tilde P(\chi) = \chi_l^{-1} P(\ln\chi_l) \propto \chi_l^{-d/z - 1}$ and for the average local susceptibility we get
\begin{equation}\label{eq:avgchi}
	\chi_{l}^{\rm (avg)} \propto \int d\chi_l~ \chi_l \tilde P(\chi_l) = \int d\chi_l~ \chi_l^{-d/z}.
\end{equation}

\begin{figure}[tbp]
	\begin{center}
		\includegraphics[width=\linewidth]{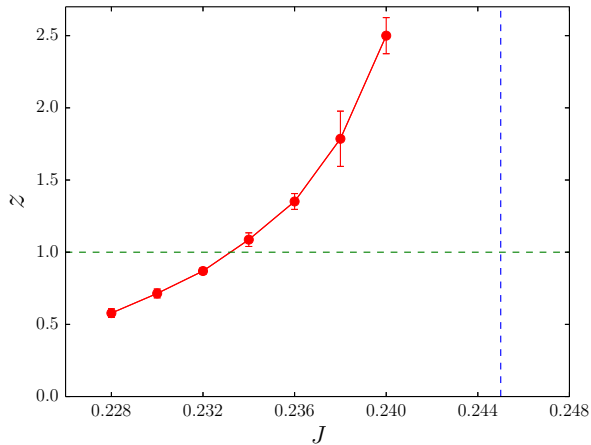}
	\end{center}
	\caption{(Color online) The dynamical exponent $z$, for different values of $J$ in the paramagnetic phase for the parameter set $(K, \Delta_K, \Delta_J) = (0.08, 0.04, 0.2)$ for our largest lattice size $L=64$. The blue vertical dashed line is the location of the induced quantum critical point. The horizontal dashed line corresponds to $z=1$.}
	\label{fig:atdynexp}
\end{figure}

In Fig.~\ref{fig:atdynexp}, we show $z$ as a function of $J$  in the paramagnetic phase. We see that the value of $z$ is larger than $1$ for a wide range of $J$ which indicates the divergence of the average local susceptibility in this region; also  $z \to \infty$ as $J \to J_c \approx 0.245$, compatible with  activated dynamical scaling at the criticality.

\section{Discussion}\label{sec:dis}

We studied the critical and off-critical properties of the quenched disorder quantum three-color Ashkin-Teller model in $(1+1)$ dimension. Through finite-size scaling analysis of the magnetic Binder cumulant at the quenched disorder induced quantum critical point, we showed that the system exhibits activated scaling. Furthermore, the calculation of the equal time correlation function showed that the rare events dominate the critical properties of the system. This results in a power law behavior of the average quantities, whereas the typical quantities exhibit a stretched exponential decay. We also calculated local susceptibility from which we extracted the dynamical critical exponent and showed the existence of Griffiths-McCoy phase away from the critical point. The overall behavior of the system is similar to the quantum spin-$\frac{1}{2}$ Ising chain, even though the pure system has a first-order transition in our case.

The critical behavior of the disorder rounded quantum first-order phase transition of the Ashkin-Teller model stands out as an example where the effect of disorder in a system is quite complex and considerable care must be exercised in analyzing quantum critical points where material disorder is inevitable.

\section{Acknowledgment}\label{sec:ackn}

We are greatly thankful to A.P. Young for important discussions. We also thank the National Science Foundation, Grant No. DMR-1004520 for support.

\end{document}